# Panoramic single-pixel imaging with megapixel resolution based on rotational subdivision

HUAN CUI,[1] JIE CAO,[1,2,*] HAOYU ZHANG,[1] CHANG ZHOU,[1] HAIFENG YAO,[2] YINGBO WANG,[4] AND QUN HAO,[1,3*]

[1] *Key Laboratory of Biomimetic Robots and Systems, Ministry of Education, Beijing Institute of Technology, Beijing, 100081, China*
[2] *Yangtze Delta Region Academy, Beijing Institute of Technology, Jiaxing 314003, China*
[3] *Changchun University of Science and Technology, Changchun, 130022, China*
[4] *School of Electronic Information and Artificial Intelligence, Shaanxi University of Science and Technology, Xi'an 710021, China*
*\*Corresponding author: ajieanyyn@163.com; qhao@bit.edu.cn*

**Abstract: Single-pixel imaging (SPI) using a single-pixel detector is an unconventional imaging method, which has great application prospects in many fields to realize high-performance imaging. In especial, the recent proposed catadioptric panoramic ghost imaging (CPGI) extends the application potential of SPI to high-performance imaging at a wide field of view (FOV) with recent growing demands. However, the resolution of CPGI is limited by hardware parameters of the digital micromirror device (DMD), which may not meet ultrahigh-resolution panoramic imaging needs that require detailed information. Therefore, to overcome the resolution limitation of CPGI, we propose a panoramic SPI based on rotational subdivision (RSPSI). The key of the proposed RSPSI is to obtain the entire panoramic scene by the rotation-scanning with a rotating mirror tilted 45°, so that one single pattern that only covers one sub-Fov with a small FOV can complete a uninterrupted modulation on the entire panoramic FOV during a once-through pattern projection. Then, based on temporal resolution subdivision, images sequence of sub-Fovs subdivided from the entire panoramic FOV can be reconstructed with pixels-level or even subpixels-level horizontal shifting adjacently. Experimental results using a proof-of-concept setup show that the panoramic image can be obtained with 10428×543 of 5,662,404 pixels, which is more than 9.6 times higher than the resolution limit of the CPGI using the same DMD. To our best knowledge, the RSPSI is the first to achieve a megapixel resolution via SPI, which can provide potential applications in fields requiring the imaging with ultrahigh-resolution and wide FOV.**

Unlike traditional imaging via a pixelated detector array, single-pixel imaging (SPI), also known as ghost imaging (GI), uses a single-pixel detector to reconstruct object information via the light-intensity correlation calculation [1-4]. With the advantages of high sensitivity, low cost, and strong anti-interference ability, SPI has great application prospects in many fields to be expected to replace traditional imaging to realize high-performance imaging, especially in special bands where it is costly or impossible to fabricate high-resolution pixelated detector arrays, such as infrared imaging [5], terahertz imaging [6], and X-ray imaging [7].

Recent growing demands for high-performance imaging at a wide field of view (FOV) or even 360° panoramic FOV [8-11], such as reconnaissance with wide FOV, panoramic situation awareness, security monitoring, medical diagnosis, pipeline detection, have led to the development of SPI with wide FOV. In 2021, the proposal of catadioptric panoramic ghost imaging (CPGI) using only a curved mirror [12,13] has taken the first step on SPI with wide FOV via extending the FOV to 360° omnidirectionally. Subsequently, super-resolution CPGI [14] and foveated CPGI [15] have improved the performance of CPGI to promote practical applications of high-performance SPI with wide FOV.

Based on the principle of CPGI, the entire omnidirectional FOV is directly covered within the illumination range of a single annular pattern through the multi-reflection on the curved mirror. This implies that the resolution of the entire panoramic image is contingent upon the resolution of the single annular pattern. However, the resolution of a single annular pattern is limited by the hardware parameters of the digital micromirror device (DMD). Taking a commonly used DMD as an example, its micromirror array is 1024×768, which means that the maximum resolution of the projected square pattern is 768 ×768. Coupled with the resolution loss caused by the annular panoramic pattern structure, the resolution limit of the entire panoramic image reconstructed by CPGI is lower than 768 ×768. This level of resolution may not meet ultrahigh-resolution panoramic imaging needs that require detailed information, making it difficult to perform advanced image processing operations that necessitate detailed visual cues.

To overcome the resolution limitation of CPGI, inspired by the mechanism of rotation-scanning panoramic imaging [16], we develop a panoramic single-pixel imaging based on rotational subdivision (RSPSI). Implementing the panoramic pattern modulation on the RSPSI uses one single pattern that can reach the maximum resolution on DMD to cover one single sub-Fov and then completes a once-through pattern modulation on the entire panoramic FOV via an uninterrupted rotation-scanning. Meanwhile, during a once-through pattern modulation, the temporal resolution [17] is introduced to subdivide the collected light intensity data to correspond to each sub-Fov, so as to reconstruct images of all sub-Fovs accurately and then stitch together into a panoramic image with entire panoramic FOV. In this way, the entire panoramic resolution can be several times higher than the resolution of the sub-Fov, while maintaining a once-through modulation via one single pattern to obtain information about the entire panoramic FOV. Note that, the panoramic resolution depends on the proportion of one single sub-Fov to the panoramic FOV, which is generally determined by the optical parameters such as the imaging lens placed in front of the DMD. Experimental results using a proof-of-concept setup of RSPSI confirm that our method can reconstruct a megapixel panoramic image with 10428 × 543 of 5,662,404 pixels, which is more than 9.6 times higher than the resolution limit of the CPGI using the same DMD.

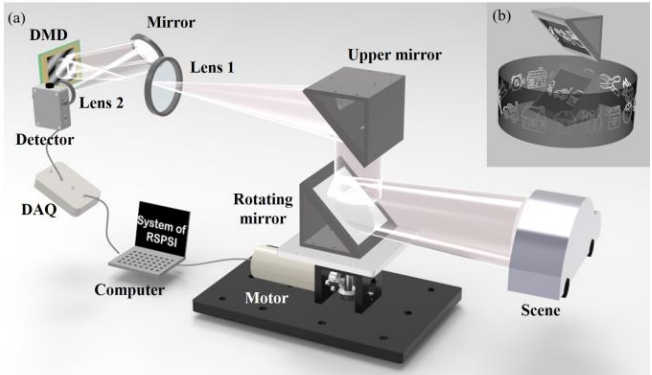

Fig.1. Experimental setup of RSPSI. (a) Structure diagram of the RSPSI. Here the rotating mirror tilted 45° is used to obtain the panoramic scene via rotation-scanning. (b) The used panoramic scene in the experiments. The panoramic scene is constructed as a hollow cylinder (height, 120mm; base diameter, 290mm) attached to 15 different objects on a black background paper, including a car, a house, a mushroom, bowling balls, etc.

The experimental setup of the proposed RSPSI is shown in Fig.1. The scene that can be self-illuminated or illuminated by a light source (a ring lamp belt is used experimentally to illuminate the cylindrical panoramic scene) is reflected by a rotating mirror tilted 45° (size, 120mm × 120mm) to an upper mirror tilted 45° (size, 90mm × 90mm). It is then focused through Lens 1 (focus length, 100mm) to image on the surface of DMD (DLP4100, 1024 × 768 micromirror array) to finish the modulation by a projected single pattern. Then, after the reflection of the Mirror and the convergence of Lens 2 (focus length, 35mm), the reflected light intensity is detected by the Detector (Thorlabs PDA36A) and collected by the DAQ (data acquisition card, PICO6404E) to transmit to the Computer (Inter i7 CPU, 16GB RAM).

The projected single pattern used in the experiments is the binary Fourier pattern [18,19], for the data efficiency of Fourier SPI, and the mathematical model is given as:

$$P(x, y, f_x, f_y, \phi) = a + b \cdot \cos(2\pi f_x x + 2\pi f_y y + \phi), \quad (1)$$

where $(x,y)$ is the spatial coordinate in one sub-Fov scene with X×Y pixels (the resolution maximum of X×Y is 1024×768), $\phi$ is the initial phase, $f_x$ and $f_y$ denote the spatial frequency in the horizontal and vertical directions respectively, $a$ is the DC term equal to the average intensity and $b$ is the contrast. Note that, the RSPSI is universal for any type of pattern, the Fourier pattern is used as an experimental example.

During a once-through pattern projection, if the rotating mirror is still, the whole setup can be used as a general SPI experimental setup. The reflected scene is only a single sub-Fov to be modulated by the projected single pattern. At this time, the light intensity detected by the DAQ is a single data without temporal resolution, which can be expressed as:

$$I(f_{x0}, f_{y0}, \phi_0) = \sum_{x=1}^{X} \sum_{y=1}^{Y} P(x, y, \phi_0, f_{x0}, f_{y0}) \bullet O(x, y), \quad (2)$$

where $O(x, y)$ represents a reflected sub-Fov scene from the entire panoramic scene.

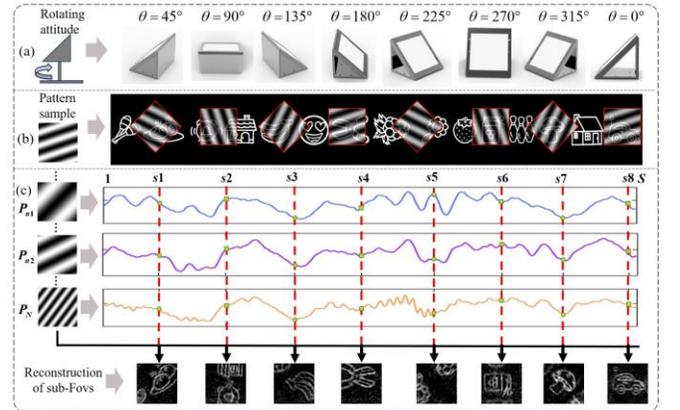

Fig.2. Imaging schematic diagram of the RSPSI. (a) Several rotating attitudes of the rotating mirror during the once-through pattern modulation. Here, the rotating attitude where the rotating mirror faces the rear imaging setup and the angle with the upper mirror is 90° is viewed as the reference attitude 0°. And the rotation-scanning is carried out clockwise. (b) Example of one single pattern modulation to sub-Fovs of the entire panoramic scene, which correspond to the reflected area from the rotating mirror with rotating attitudes on (a). (c) Reconstruction of sub-Fovs. Here, with a sampling ratio of 0.15%, the image of each sub-Fov with 768×768 pixels is reconstructed via the four-step phase-shifting Fourier SPI with the measurements number of $N$ =1766. Note that, $s1$-$s8$ correspond to the collected light intensity values sequences of the reflected sub-Fov scenes from the rotating mirror with different rotating attitudes on (a) to reconstruct the images having an image rotation effect.

However, during the same once-through pattern projection, if the rotating mirror rotates 360° uniformly by a servo motor (ASSUN, AM-BL45100AE) drive control synchronously, then the whole setup plays the role of the RSPSI to reflect the entire panoramic scene through the uninterrupted rotation-scanning. In this way, one single pattern can modulate uninterruptedly the entire panoramic FOV during the once-through projection. As shown in Fig.2, under different rotating attitudes of the rotating mirror, the single pattern covers different sub-Fovs corresponding to the reflected areas from the panoramic scene. At this time, the light intensity detected by the DAQ is a 1-D data sequence with temporal resolution, which can be expressed as:

$$\begin{bmatrix} I_1(f_{x0},f_{y0},\phi_0) \\ I_2(f_{x0},f_{y0},\phi_0) \\ \vdots \\ I_s(f_{x0},f_{y0},\phi_0) \\ \vdots \\ I_S(f_{x0},f_{y0},\phi_0) \end{bmatrix} = \sum_{x=1}^{X}\sum_{y=1}^{Y} P\begin{pmatrix} x,y,\phi_0 \\ f_{x0},f_{y0} \end{pmatrix} \bullet \begin{bmatrix} O_1(x,y) \\ O_2(x,y) \\ \vdots \\ O_s(x,y) \\ \vdots \\ O_S(x,y) \end{bmatrix}, \quad (3)$$

where $S$ is called the temporal subdivision factor, which represents the subdivision capability of temporal resolution. The larger $S$ is, the higher the sampling frequency of the DAQ, the denser sub-Fovs, and the smaller the shifting of adjacent sub-Fovs. $O_s(x,y)$ represents the $s$th subdivided sub-Fov scene of the panoramic FOV and $I_s$ denotes the collected light intensity value reflected from the $s$th subdivided sub-Fov scene after the modulation by one single pattern.

Then, after completing the required $N$-times projection modulation, based on the temporal subdivision characteristic, the collected light intensity data sequence with the size of $S \times N$ can be registered and sorted out to $S$ light intensity data sequences with the size of $1 \times N$ about $S$ subdivided sub-Fov scenes, and then images of $S$ subdivided sub-Fov scenes can be reconstructed. Using the reconstruction algorithm of the four-step phase-shifting Fourier SPI, the image of the $s$th subdivided sub-Fov scene $O'_s(x,y)$ can be obtained as:

$$O'_s(x,y) = F^{-1}\left\{ \begin{array}{l} \left(I_s(f_x,f_y,0) - I_s(f_x,f_y,\pi)\right) \\ + j\left(I_s(f_x,f_y,\frac{\pi}{2}) - I_s(f_x,f_y,\frac{3\pi}{2})\right) \end{array} \right\}, \quad (4)$$

where $j$ denotes the imaginary unit and $F^{-1}$ denotes the inverse Fourier transform. The number of pattern projections, that is, the measurements number of each subdivide sub-Fov, $N$, depends on the resolution of the single pattern, the required sampling rate, and the used SPI reconstruction algorithm, which is independent of the resolution of the entire panoramic FOV and the number of subdivided sub-Fovs.

It is worth noting that, at the rotation-scanning, the normal direction of the rotating mirror rotates following different rotating attitudes. Then, the reflected sub-Fov scene rotates synchronously but is modulated by irrotational patterns projected on the stationary DMD, which would generate an image rotation effect on the reconstructed images of sub-Fovs, as shown in Fig.2. The image rotation effect makes it impossible to directly carry out the registration and stitching reconstruction of the horizontal panoramic FOV. Therefore, to adapt to the human visual system and complete the horizontal stitching of the entire panoramic FOV, the affine transformation matrix is used to offset the image rotation effect on the reconstructed images of subdivided sub-Fovs. According to the reflection principle, the image rotation angle is consistent with the corresponding rotation attitude angle $\theta$ of the rotating mirror. Therefore, given the center point $(x_c, y_c)$, rotating the angle $\theta$ counterclockwise, the coordinate of the $s$th sub-FOV rectified image $RO'_s(x, y)$ can be expressed as:

$$\begin{bmatrix} x' - x_c \\ y' - y_c \end{bmatrix} = \begin{bmatrix} \cos\theta & -\sin\theta \\ \sin\theta & \cos\theta \end{bmatrix} \cdot \begin{bmatrix} x - x_c \\ y - y_c \end{bmatrix}. \quad (5)$$

Besides, the bilinear interpolation is used to resample the rectified image to ensure that the effective pixels remain unchanged to avoid imaging quality degradation during the rotation transformation. However, in this way, the invalid pixels of the rectified image, such as the black-filled area shown in Figure 3 (a), make the images of adjacent sub-Fovs discontinuous, which can cause the stitching mismatch. To ensure that the adjacent sub-Fovs are horizontally coherent and matchable, according to the rotation symmetry, the inner square area of the tangent circle via central cropping is selected as the effective region of sub-Fov for stitching, and the unilateral size of the cropped image is given as:

$$M = \min(X,Y)/\sqrt{2}. \quad (6)$$

To avoid resolution redundancy sampling of the sub-Fov, the shape of a single pattern is set as a square with $X \times X$ pixels, where the maximum resolution of a single pattern is 768×768 pixels experimentally and the corresponding resolution of the cropped image of a sub-Fov is 543×543 pixels. As the offset error caused by the precision error of the mechanical assembly and the non-uniformity of the rotation-scanning speed is ignored, adjacent cropped images of sub-Fovs only shift horizontally in theory, and the shifting amount is only related to the temporal subdivision factor $S$. Given the horizontal FOV angle of one pixel is pf, then $S = 2\pi/(h \times pf)$, where h represents the horizontal shifting amount between adjacent sub-Fovs. With a fixed projection time, the minimum limit of h depends on the response speed of the Detector and the highest sampling frequency of the DAQ. Therefore, a pixels-level or even subpixels-level horizontal space matching mode can be established from the cropped image sequences of sub-Fovs with $h$-pixel horizontal shifting adjacently. The panoramic image can be reconstructed by matching the position of each cropped sub-Fov image in a panoramic resolution grid with $F \times M$ pixels ($F=2\pi/pf$, which is the total horizontal pixels number of the cropped panoramic image; $M$ is the total vertical pixels number of the cropped panoramic image), in which the overlapping parts of adjacent sub-Fovs are smoothed by weighted average fusion. Further, when $h<1$, the subpixel horizontal shifting between adjacent sub-Fovs can provide subpixel-level complementary detail information to further improve the resolution of the panoramic FOV and the imaging quality of the sub-Fovs by super-resolution algorithms [20].

Using the experimental setup of the RSPSI, with the maximum resolution, the measured FOV angle of one pixel $pf$ is known as about 6.0253e-4 rad. After matching and stitching the cropped sub-Fov images sequence with 543×543 pixels and one-pixel horizontal shifting adjacently, the cropped panoramic image with 10428×543 pixels is obtained, as shown in Fig.3 (c). To our best knowledge, it is the first to achieve a megapixel resolution via SPI, which is 19.2 times higher than the resolution of one single cropped sub-Fov image, as well as 7.2 times higher than the maximum resolution limited by the used DMD. Furthermore, $pf$ can be smaller by the further design of optical components, such as a telephoto lens, to obtain a higher resolution on the entire panoramic FOV.

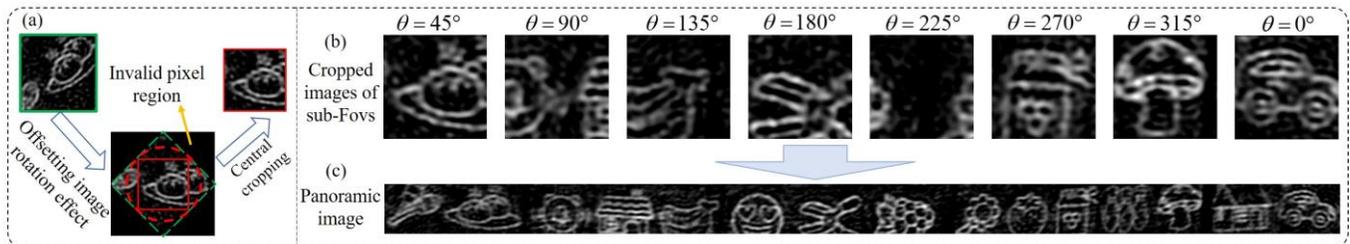

Fig.3. Experimental results of RSPSI. (a) Example of the rectification of one sub-Fov for stitching. By offsetting the image rotation effect, the original reconstructed image of one sub-Fov scene is rectified, and then the central cropping makes the adjacent sub-Fovs continuous and matchable. (b) Cropped images of sub-Fovs with 543×543 pixels corresponding to different rotating attitudes of the rotating mirror. (c) Panoramic image with 10428×543 pixels by matching and stitching the cropped sub-Fov images sequence of (b). Video 1 shows the complete imaging of RSPSI.

We acknowledge that acquisition comes at a cost, the RSPSI is limited in real-time at its current stage, despite achieving an ultrahigh resolution. Even though the rotating mirror has the advantages of small size, high stability, simple structure, etc., due to the limitation of the mechanical manufacturing process and motor speed, the uniform rotation-scanning speed is difficult to reach or exceed the maximum refresh rate of DMD (the refresh rate of 22kHz corresponds to the rotation-scanning speed of 1.32e6 rpm), which will affect the imaging efficiency of the RSPSI. Thus, our future work is to find effective methods to improve or compensate the imaging efficiency of the RSPSI, so as to break through the bottleneck of SPI with panoramic FOV, high resolution, and real-time.

In conclusion, we propose a high-resolution panoramic SPI method that can achieve megapixels or even higher. The key of the proposed RSPSI is to obtain the entire panoramic scene by the rotation-scanning with a rotating mirror tilted 45°, so that one single pattern that only covers one sub-Fov with a small FOV can complete the uninterrupted modulation on the entire panoramic FOV during a once-through projection. Then, based on temporal resolution subdivision, the detected light intensity values during the once-through pattern modulation are finely allocated to reconstruct images of subdivided sub-Fovs with maximum resolution up to the DMD hardware limit. In addition, due to the fast response speed of the single-pixel detector and the high acquisition frequency of the DAQ, the shifting between adjacent sub-Fovs can reach pixels-level or even subpixels-level, so that a pixels-level or even subpixels-level horizontal space matching mode can be established, and the high-resolution panoramic image can be stitched more accurately and effectively. The proof-of-concept experimental results show that except for the clipped areas to offset the image rotation effect caused by the rotation-scanning of the rotating mirror tilted 45°, the panoramic image is obtained with 10428×543 pixels as each sub-Fov reaches the maximum resolution limit, which is more than 9.6 times higher than the resolution limit of the CPGI using the same DMD. The proposed RSPSI provides potential applications in fields requiring imaging with ultrahigh-resolution and wide FOV. Furthermore, advanced methods for improving resolution and imaging efficiency can be further combined to achieve higher performance and accelerate the practical process of the RSPSI.